\documentstyle{l-aa}
\font\sixrm=cmr6
\newcommand{\magpt}[2]{\mbox{$\rm #1\hspace{-0.25em}\stackrel{m}{.}
      \hspace{-1.0mm}#2$}}                             
\def\bsec{\hbox{$.\!\!{\arcsec}$}}
\def\grad{\hbox{$^\circ$}}
\def\bmin{\hbox{$^\prime$}}
\def\std{\hbox{$^{\rm h}$}}
\def\min{\hbox{$^{\rm m}$}}
\def\dsec{\hbox{$.\!\!^{\rm s}$}}
\newcommand{\RA}[4]{#1\std #2\min #3\dsec#4} 
\newcommand{\DEC}[4]{$#1$\grad $#2$\bmin #3\bsec#4}      
\newcommand\ion[2]{\hbox{#1\,{\sixrm #2}}}
\newcommand\Ebv{$ {\rm E_{B-V}}$}
\newcommand\teff{$ {\rm T_{eff}}$}
\newcommand\logt{$\log {\rm T_{eff}}$}
\newcommand\logg{$\log {\rm g}$}
\newcommand\logl{$\log {\rm L/L_{\odot}}$}
\newcommand\loghe{${\rm \log{\frac{n_{He}}{n_{H}}}}$}
\newcommand{\Msolar}{\mbox{\,$\rm M_{\odot}$}}        
\begin{document}
\thesaurus{5(08.16.4 - 10.07.3 NGC 2808 - 10.07.3 NGC 6121 - 
10.07.3 NGC 6723 - 10.07.3 NGC 6752)}
\title{Hot UV bright stars in globular clusters
\thanks{Based on observations collected at the European Southern Observatory
(ESO N$^\circ$ 57.E-0101)}}
\author{S. Moehler
\inst{1}$^,$\inst{2} \and W. Landsman \inst{3}
 \and R. Napiwotzki \inst{1} }
\offprints{S. Moehler}
\institute {Dr. Remeis-Sternwarte, Astronomisches Institut der Universit\"at 
Erlangen-N\"urnberg, Sternwartstr. 7, 96049 Bamberg, Germany
\and Space Telescope Science Institute, 3700 San Martin Drive, Baltimore,
MD 21218, USA
\and Raytheon STX, NASA/GSFC, Greenbelt, MD 20770, USA}
\date{}
\maketitle
\keywords{Stars: Post-AGB -- globular clusters individual: NGC 2808 -- NGC 
6121 -- NGC 6723 -- NGC 6752}

\begin{abstract} 

We have obtained medium-resolution spectra of seven UV-bright stars discovered
on images of four southern globular clusters obtained with the  Ultraviolet
Imaging Telescope (UIT).      Effective temperatures, surface gravities and
helium abundances are derived from LTE and non-LTE model atmosphere fits.   
Three of the stars have sdO spectra,  including M4-Y453 (\teff\ = 58800~K,
\logg\ = 5.15), NGC 6723-III60 (\teff = 40600~K, \logg\ = 4.46) and 
NGC~6752-B2004 (\teff\ = 37000~K, \logg\ = 5.25). All seven stars lie along
either post-extended horizontal branch (EHB) or post-early AGB evolutionary
tracks.     The post-early AGB stars show solar helium abundances, while the
post-EHB stars are helium deficient, similar to their EHB progenitors.

\end{abstract}

\section{Introduction}

Ultraviolet images of globular clusters are often dominated by  one or two 
hot, luminous, ``UV-bright'' stars.  The most luminous of these stars are
believed to be post-asymptotic giant branch (post-AGB) stars, which go through
a luminous UV-bright phase as they leave the AGB and move rapidly across the HR
diagram toward their final white dwarf state.    Despite their short lifetimes
($\sim 10^{5}$ yrs), hot post-AGB stars can dominate the total ultraviolet flux
of an  old stellar population.    In particular, hot post-AGB stars
probably make a  significant (although not the dominant) contribution  to the
UV-upturn observed in elliptical galaxies (Brown et al.\ \cite{brown97}).    
However, a large uncertainty exists in modeling the contribution of hot
post-AGB stars to the integrated spectrum of an old stellar population, due to
the strong dependence of the post-AGB luminosity and lifetime on the core mass,
which in turn depends  on when the stars leave the AGB (Charlot et al.\
\cite{charris96}).  Also the previous  mass loss on the red giant branch
(RGB) plays an important r\^ole here,  since it determines the fate of a star
during and after the horizontal branch  stage: Stars with very low envelope
masses settle along the extended  horizontal branch (EHB) and evolve from there
directly to the white dwarf  stage,  whereas stars with envelope masses of more
than 0.02~\Msolar  will at least partly ascend the AGB.  A further uncertainty
arises because theoretical post-AGB tracks have been only minimally tested for
old, low-mass stars.    

The last census of hot post-AGB stars in globular clusters was published by de
Boer (\cite{debo87}), but this list  is certainly incomplete.   The detection 
of hot post-AGB stars in optical  color-magnitude diagrams (CMD's) is
limited by selection effects due to crowding in the cluster cores and to the
large bolometric corrections for these hot stars.   More complete searches are
possible for hot post-AGB stars in planetary nebulae, for example, by
using \ion{O}{III} imaging.   However, only four planetary nebulae (PNe) were
discovered in a recent survey of 133  globular clusters (Jacoby et al.\
\cite{jamo97}), of which two were previously known (K648 in M~15, and IRAS
18333-2357 in M~22).        Jacoby et al. expected to find 16 planetary nebulae
in their  sample, on  the basis of the planetary nebula luminosity function for
metal-poor populations.    The origin of this discrepancy is not yet
understood, but we mention two possible contributing factors.   First, the 
\ion{O}{III} search of Jacoby et al. may have missed some old, faint  planetary
nebulae.     Second, Jacoby et al. derive  the number of expected PNe from the
total cluster luminosity, assuming  that all stars in a globular cluster will
eventually go through the AGB phase.    But in a cluster such as NGC 6752,
about 30\% of the HB population consists of EHB stars  (with \teff\ $>$ 
20,000~K),  which are predicted to evolve into white dwarfs without ever passing
through the thermally pulsing AGB phase. 
The exact fraction of stars which follow such evolutionary will depend on 
the poorly known mass loss rates during the HB and early-AGB phases.

While globular clusters with a populous EHB are expected to be deficient in
post-AGB stars, they should show a substantial population of less luminous 
(1.8 $<$ \logl\ $<$ 3) UV-bright stars, which can be either  post-EHB  stars or
post-early AGB stars. The population of  post-EHB stars  is expected to be
about 15--20 \% of the population of EHB stars (Dorman et al. \cite{dorm93}).  
The post-early AGB population arises from hot HB stars with sufficient envelope
mass to return to the AGB, but which peel off  the AGB prior to the thermally
pulsing phase (Dorman et al. \cite{dorm93}). 

During the two flights of the {\em ASTRO} observatory in 1990 and 1995, the
Ultraviolet Imaging Telescope (UIT, Stecher et al.\ \cite{stech97}) was used to
obtain ultraviolet ($\sim 1600$~\AA) images of 14 globular clusters.   The
solar-blind detectors on UIT suppress the cool star population, which allows
UV-bright stars to be detected into the cluster cores, and the $40'$ field of
view of UIT is large enough to image the entire population of most of the
observed clusters.     Thus, the UIT images provide a complete census of the
hot UV-bright stars in the observed clusters.   We have begun a program to
obtain spectra of all the UV-bright stars found on the UIT images, in order to
derive effective temperatures and gravities for the complete sample, for
comparison with evolutionary tracks.       Several of the UV-bright stars found
on the UIT images, such as ROB 162 in NGC~6397, Barnard 29 in M~13, and  vZ
1128 in M~3, were previously known and are well-studied.   Other UIT stars are
too close to the cluster cores for ground-based spectroscopy, and will require
HST observations for further study.    In this paper, we report on spectroscopy
of those UIT UV-bright stars accessible for ground-based observations from the
southern hemisphere.

\section{Observations}

\begin{table*}
\begin{tabular}{|ll|ll|ll|cc|}
\hline
Cluster & Star & $\alpha_{2000}$ & $\delta_{2000}$ & V & B$-$V & slit & 
seeing \\
 & & & & & & orientation & \\
    &   &   &   & [mag] & [mag] & & [\arcsec] \\
\hline
NGC 2808 & C2946$^1$ & \RA{09}{12}{22}{36} & \DEC{$-$64}{52}{37}{3} & 17.63 
& +0.09 & EW/NS & 1.0 -- 1.9\\
    & C2947$^1$ & \RA{09}{12}{22}{86} & \DEC{$-$64}{52}{36}{7} & 17.10 & +0.21 
& EW/NS & 1.0 -- 1.9\\
    & C4594$^1$ & \RA{09}{12}{01}{98} & \DEC{$-$64}{47}{35}{1} & 16.36 & +0.03
& NS &  1.6 -- 1.9\\
 & & & & & & & \\
NGC 6121 & Y453$^2$ & \RA{16}{23}{22}{25} & \DEC{$-$26}{28}{02}{4} & 15.86 & 
+0.00 & NS & 1.0 -- 1.4\\
 & & & & & & & \\
NGC 6723 & III-60$^3$ & \RA{18}{59}{29}{0} & \DEC{$-$36}{40}{49}{0} & 15.61  &
$-$0.25 & NS & 2.6 -- 3.9\\
         & IV-9$^3$ & \RA{18}{59}{24}{1} & \DEC{$-$36}{37}{56}{0} & 14.64$^4$  &
 $-$0.17$^4$ & EW & 1.5 -- 1.6\\
 & & & & & & & \\
NGC 6752 & B2004$^5$ & \RA{19}{11}{04}{9} & \DEC{$-$59}{57}{47}{0} & 16.42 & 
$-$0.31 & EW/NS & 1.2 -- 1.7\\
\hline
\end{tabular}
\begin{tabular}{l}
$^1$ Ferraro et al.\ (\cite{ferr90}) (This photometry has been adjusted
0.1 mag fainter, as reported by Sosin et al., \cite{sos97})\\
$^2$ Cudworth \& Rees (\cite{cud90}) \\
$^3$ Menzies (\cite{menz74})\\
$^4$ L.K. Fullton (priv. comm.)\\
$^5$ Buonanno et al.\ (\cite{buca86})\\
\end{tabular}
\caption[]{List of observed stars and observing parameters}
\end{table*}

In May 1996 we took medium resolution spectra of UV bright  stars in southern
globular clusters using EFOSC2 at the 2.2m MPI/ESO  telescope (Table 1).  We
used grism \#3 (100~\AA/mm, 3800 -- 5000~\AA) and a slit  width of 1\arcsec,
resulting in a resolution of 6.7~\AA . The instrument was equipped with a
Thompson CCD (19$\mu$m pixel size,  1024 $\times$ 1024 pixels, gain
2.1~e$^-$/ADU, read-out-noise 4.3~e$^-$). For calibration purposes we always
observed 10 bias frames each night and 5-10 flat-fields with a mean exposure
level of about 10000 ADU each. For observations longer than 30 minutes we took
wavelength calibration frames before and after the object's observation. Since
the Ar light is rather faint, we took HeAr frames at the beginning of the night
and only He during the night. The seeing values and slit orientations are
listed in Table 1.

\section{Data Reduction}

\begin{figure}
\vspace{11cm}
\includegraphics{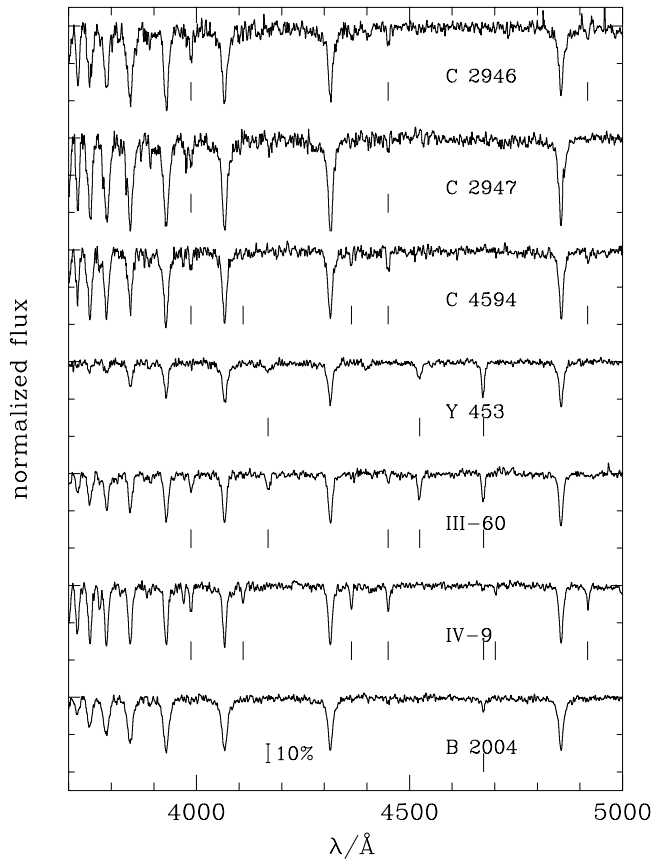}
\caption[]{The normalized spectra of the UV bright stars. The \ion{He}{I} 
and \ion{He}{II} absorption lines are marked.}
\end{figure}

We averaged the bias frames over all nights of the run, as they showed no
deviations above the 1\% level. For the bias correction we added the difference
in the mean overscans of the science and the bias frame to the mean bias frame,
which was then  subtracted from the science frame. We did not correct for dark
current, which was less than 3~cts/hr/pix and showed no discernible structure.
The flat fields were averaged separately for each night and normalized by
dividing them 
by a smoothed average along the spatial axis. The flat fields showed
night-to-night variations of less than 1\% between the first and second night
and again between the third and fourth night. We therefore averaged the 
flat-fields of the first and second night and those of the third and fourth 
night. \\ \indent For the wavelength calibration we used a 3$^{\rm rd}$ order
polynomial to fit the dispersion relation for the  HeAr frames. We used 15
lines,  avoiding blended lines. The fits then yielded mean residuals of less
than 0.2~\AA. From the He frames obtained during  the night we derived offsets
relative to the HeAr frames by cross correlating  the spectra, which were then
used to adjust the zero-points of the  dispersion relations.\\  \indent We
rebinned the frames two-dimensionally to constant wavelength steps. Before the
sky fit the frames were median filtered along the spatial axis to  remove
cosmic rays in the background. To determine the sky background we had  to find
regions not containing any stellar spectra, which were sometimes  distant
from the object's location.  Nevertheless the flat field correction and
wavelength calibration turned out to be good enough that a constant fit to the
spatial distribution of the sky light allowed us  to subtract the sky
background at the object's position with sufficient  accuracy in the less 
crowded regions of the globular clusters, i.e. we do not see any
absorption lines caused by the predominantly red stars of the clusters or by
moon light. For those stars that lay closer to the cluster center the
background light of the cluster showed a gradient across the frame and we
therefore used a linear fit for the spatial distribution of the background
light. The fitted sky background was then subtracted from the unsmoothed frame
and the spectra were extracted using Horne's algorithm (Horne, \cite{horn86})
as implemented in MIDAS.\\ The normalized spectra are plotted in Figure 1.
Spectra of the UV-bright stars ROA 542 and ROA 3596 in $\omega$ Cen obtained
during this run will be reported in a separate paper (Landsman et al.\
\cite{land98}).

\section{Spectroscopic analyses}

To derive effective temperatures, surface gravities and helium abundances  we
fitted the observed Balmer and helium lines with appropriate stellar  model
atmospheres. Beforehand we corrected the spectra for radial velocity  shifts,
derived from the positions of the Balmer and helium lines. The  resulting
heliocentric velocities are listed in Table 2, together with the  physical
parameters of the stars. To establish the best fit we used the routines
developed by Bergeron et al.  (\cite{besa92}) and Saffer et al.
(\cite{saff94}), which employ a $\chi^2$  test.

\subsection{Subdwarf O stars} Those stars which show \ion{He}{II} lines in
their spectra (Y453 in NGC~6121, III-60 in NGC~6723, and B2004 in NGC~6752) are
classified as sdO stars.  To analyse these stars correctly it is necessary to
take non-LTE effects into  account. NLTE model atmospheres were calculated with
the code developed by  Werner (\cite{wern86}). The basic assumptions are
those of static, plane-parallel atmospheres in hydrostatic and radiative
equilibrium. In contrast to the ATLAS9 atmospheres used to analyse the  cooler
programme stars, we relax the assumption of local thermal equilibrium  (LTE)
and solve the detailed statistical equilibrium instead. The  accelerated
lambda iteration (ALI) method is used to solve the non-linear  system of
equations as described in Werner (\cite{wern86}). 

To keep the computational effort within reasonable limits we neglected the
influence of heavy elements and assumed a  mixture of hydrogen and helium
only. Elaborate hydrogen and helium model atoms were used for this purpose and
pressure dissolution of the higher levels is included according the  Hummer \&
Mihalas (\cite{humi88}) occupation probability formalism. Details are given in
Napiwotzki (\cite{napi97}). Our NLTE model grid covers the temperature range
$27000 < $ \teff\ $< 70000$\,K (stepsize increasing with \teff\ from 2000\,K to
5000\,K) and gravity range $3.5 <$ \logg $< 7.0$ (stepsize 0.25\,dex) with the
helium abundance varying from \loghe\  = $-3.0$ to $+0.5$ in 0.5\,dex steps.
The fits for the parameters given in Table 2 are shown in Fig. 2.

\begin{figure*}
\vspace{7cm}
\includegraphics{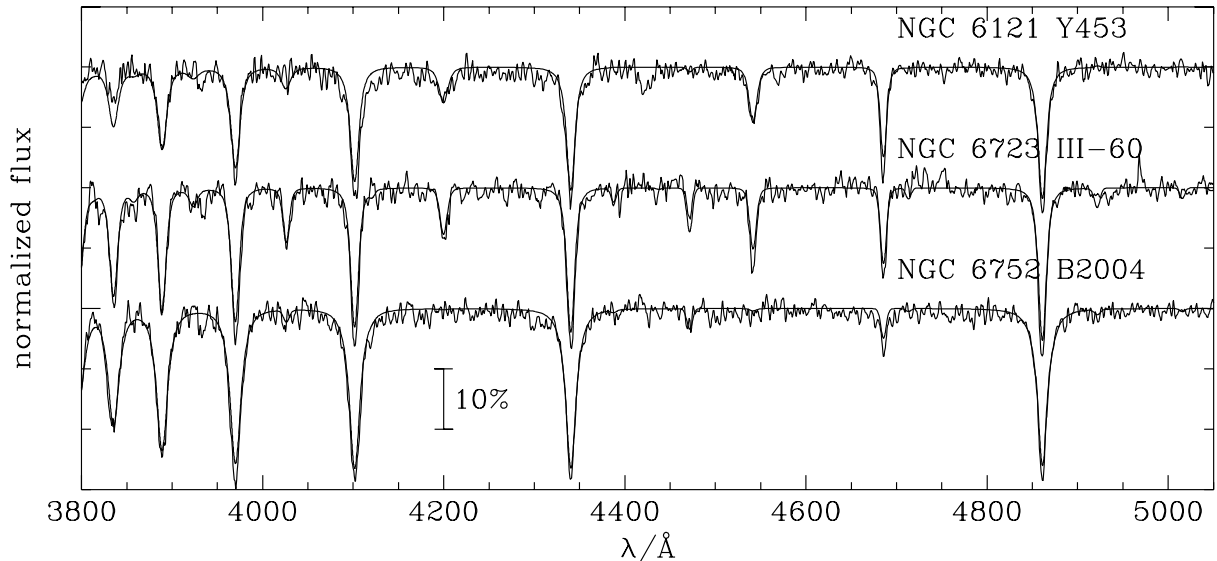}
\caption[]{The model atmosphere fits for the sdO stars. }
\end{figure*}

\subsection{Cooler UV bright stars}

The remaining stars showed only Balmer and \ion{He}{I} lines in their spectra.  
For these stars LTE model  atmospheres are sufficient to derive physical
parameters (Napiwotzki \cite{napi97}). We computed model atmospheres using
ATLAS9 (Kurucz 1991, priv. comm.) and  used the  LINFOR program (developed
originally by Holweger, Steffen, and  Steenbock at Kiel university) to compute
a grid of  theoretical spectra, which include the Balmer lines H$_\alpha$ to
H$_{22}$ and \ion{He}{I} lines. The grid covered the range 10000 \ldots 27500~K
in \teff, 2.5 \ldots 5.0 in  \logg\ and $-$2.0 \ldots $-$0.3 in \loghe\ at  
metallicities of $-$2 and $-$1 (Lemke, priv. comm.).

We fitted the Balmer lines from H$_\beta$ to  H$_{12}$ (excluding H$_9$ 
because of the interstellar \ion{Ca}{II} H line)  and (if present) the
\ion{He}{I} lines $\lambda\lambda$ 4026~\AA , 4388~\AA ,  4472~\AA, 4922~\AA ,
and 5016~\AA . Fitting the spectra of the three stars in  NGC~2808 and IV-9 in
NGC~6723 with models  for both metallicities resulted in temperature
differences of about 500~K (the temperature being higher for the lower
metallicity). The surface  gravities and  helium abundances were not affected.
We decided to keep the values  determined for a metallicity of $-$1.0, since
this value is closer to the  actual metallicities of these two clusters. The
results of the spectroscopic analyses are given in Table 2.

\begin{table*}
\begin{tabular}{|llll|rrrrr|ccc|}
\hline
Cluster & [Fe/H] & v$_{\rm hel,Cl.}$ & Star & \teff & \logg & 
${\rm \log{\frac{n_{He}}{n_{H}}}}$ & v$_{\rm hel}$ & M & status 
 & ${\rm {\log{\left(\frac{L}{L_\odot}\right)_{UV}}}}$ 
 & ${\rm {\log{\left(\frac{L}{L_\odot}\right)_{V}}}}$ \\
   &   & [km/s] & &  [K]  &       &                  & [km/s] & 
[M$_\odot$] & & & \\
\hline
NGC 2808 & $-$1.37 & +94 & C2946 & 22700 & 4.48 & $-$1.72 & +93 & 0.46 & 
post-EHB & & +1.99 \\ 
    & & & C2947 & 15100 & 3.82 & $-$1.21 & +134 & 0.32 & post-EHB & & +1.78 \\
    & & & C4594 & 22700 & 4.06 & $-$1.57 & +89 & 0.55 & post-EHB & +2.44 & 
 +2.41\\
 & & & & & & & & & & & \\
NGC 6121 & $-$1.20 & +70 & Y453   & 58800 & 5.15 & $-$0.98 & +31 & 0.16 & 
post-EAGB & +2.61 & +2.54 \\ 
 & & & & & & & & & & & \\
NGC 6723 & $-$1.12 & $-$95 & III-60 & 40600 & 4.46 & $-$1.03 & $-$109 & 0.49  
& post-EAGB & +2.92 & +3.06 \\
 & & & IV-9 & 20600 & 3.34 & $-$0.83 & $-$52  & 0.29 & post-EAGB & +2.82 & 
+2.79 \\
 & & & & & & & & & & & \\
NGC 6752 & $-$1.55 & $-$25 & B2004 & 37000 & 5.25 & $-$2.39 & 0 & 0.32 
& post-EHB & +1.94 & +1.94 \\ 
\hline
\end{tabular}
\caption[]{List of observed stars and their atmospheric parameters. The 
metallicities and radial velocities for the clusters 
are taken from the May 1997 tabulation of Harris (\cite{harris96}).}
\end{table*}

\begin{figure} \vspace{7.5cm} \includegraphics{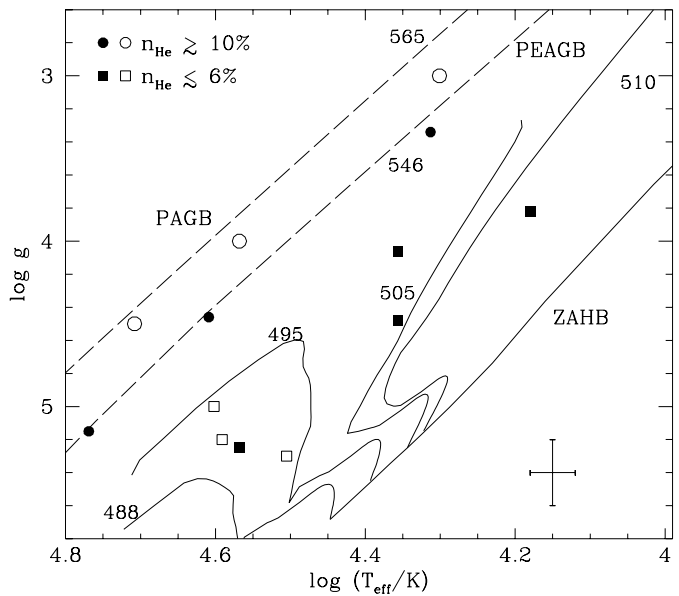} \caption[]{The atmospheric parameters of the
UV bright stars compared to  evolutionary tracks. The solid lines mark the ZAHB
and post-HB evolutionary tracks for [Fe/H] = $-$1.48  (labeled with the mass of
the tracks  in units of 10$^{-3}$\Msolar, Dorman et al., \cite{dorm93}).  The
dashed lines give post-AGB (0.565~\Msolar) and post-early AGB (0.546~\Msolar)
tracks from Sch\"onberner (\cite{scho83}), also labeled with the mass of the
tracks in units of 10$^{-3}$\Msolar. The filled symbols are from this paper,
the open symbols are taken from the literature (Conlon et al., \cite{codu94};
Heber \&  Kudritzki, \cite{heku86}; Heber et al., \cite{hedr93}; Moehler et
al.,  \cite{mohe97}). } \end{figure}

\section{Luminosities and masses}

To compute the luminosities, we require the cluster distances and reddenings.
These values are taken from the May 1997 tabulation of Harris (\cite{harris96})
except that the distance and reddening toward M~4 (d = 1.73 kpc, \Ebv\ = 0.35)
is taken from the HST study of Richer et al. (\cite{rich97}).    We also use
the  nonstandard value of 3.8 for R$_V$ [=A$_V$/\Ebv ] toward M~4, as suggested
recently by several   authors (Peterson et al. \cite{pe95}; Richer et al.\
\cite{rich97}), along with the ISM parameterization of Cardelli et al.
(\cite{ccm89}). \\

Luminosities were determined by first using the distance and reddening to
convert the V or UIT magnitude to an absolute magnitude, and then applying a
bolometric correction using the \teff\ derived from the model atmosphere fit.
Because these hot stars have large bolometric corrections, the luminosities
derived using the V magnitude are very sensitive to the value of \teff\
($\Delta \log$L $\sim 3 \Delta \log$~\teff). The UIT magnitude has a much
smaller bolometric correction,  but typically has poorer photometric precision,
and is more sensitive to reddening.    Therefore, we have computed luminosities
using both methods. As can be seen from Table~2 both values  agree rather well.   
The uncertainty in the derived luminosities  ranges from 0.12 dex for the
cooler stars, to 0.16 dex to NGC 6121-Y453, assuming uncertainties of
\magpt{0}{1} in V, \magpt{0}{15} in the UIT flux, \magpt{0}{02} in \Ebv,  0.035
dex in \teff, and 10\% in the cluster distances.

Knowing effective temperatures, surface gravities and absolute magnitudes  we
can derive the masses of the stars as described in Moehler et al.
(\cite{mohe97}). Assuming errors of \magpt{0}{1} and  \magpt{0}{13}  in
the observed and theoretical V magnitudes, respectively, and 0.15 dex in 
\logg\ we arrive at an error for log (M/\Msolar) of 0.18 dex or about 50\%. 
For a detailed discussion of error sources see Moehler et al. (\cite{mohe97}).
The results are listed in Table~2.

\section{Discussion}

The derived effective temperatures and gravities of the target stars are
plotted in Figure 3, along with  ZAHB and post-HB evolutionary tracks for 
[Fe/H] = $-$1.48 from Dorman et al. (\cite{dorm93}), and post-AGB
(0.565~\Msolar) and post-early AGB (0.546~\Msolar) tracks from Sch\"onberner
(\cite{scho83}).  The stars NGC~6121-Y453, NGC~6723-III60, and NGC~6723-IV9
appear to fit the  post-early AGB track, while the remaining four targets are
consistent with post-EHB evolutionary tracks.  In agreement with this scenario,
the three post-early AGB stars have approximately solar helium abundances,
while the post-EHB stars have subsolar helium abundances.    The latter stars
are expected to have subsolar helium abundances because they are direct
descendants of EHB stars, which are known to show helium deficiencies (Moehler
et al. \cite{mohe97}), most likely due to diffusion processes.     The
post-early AGB stars, on the other hand, have evolved off the AGB,  where the
convective atmosphere is expected to eliminate any previous abundance
depletions  caused by diffusion. Curiously, no helium-rich (He/H $\ge$ 1)
sdO stars have yet been found in a  globular cluster, although such stars
dominate the field sdO population (Lemke et al. \cite{lehe98}).  

As expected, the two clusters with a populous EHB (NGC~2808 and NGC~6752) have
post-EHB stars but no post-AGB stars. The clusters NGC~6723 and M~4, on  the
other hand,  do not have an EHB population, although they do have stars
blueward of the RR Lyrae gap (which are  potential progenitors of post-early
AGB stars). The lack of true post-AGB stars may be understood from the 
different lifetimes: The lifetime of  Sch\"onberner's post-early AGB track is
about 10 times longer than his lowest mass post-AGB track.  Thus, even if only
a small fraction of stars follow  post-early AGB tracks, those stars may be
more numerous than true  post-AGB stars. Due to their relatively long
lifetime, post-early AGB stars are unlikely to be observed as central stars of
planetary nebulae (CSPNe) since any  nebulosity is probably dispersed before
the central star is hot enough to  ionize it.   Additional detail on the
individual stars is given below:

\subsection{NGC 2808}

All three stars in NGC~2808 analysed in this paper are likely post-EHB stars.
(Unfortunately, the best post-AGB candidate stars on the UIT image of 
NGC~2808 are too close to the cluster center to allow spectroscopy from the
ground.) Although C2946 and C2947 could be separated in the long-slit optical
spectra, they are too close together to estimate individual UV fluxes from the
UIT image, and thus there is no UV luminosity determination in Table 2. Due to
the well-populated EHB of NGC~2808 (Sosin et al., \cite{sos97}), a large number
of post-EHB stars are expected.   

From their three-colour WFPC2 photometry of NGC 2808, Sosin et al.\
(\cite{sos97}) find a larger distance modulus  [(m-M)$_0$ = 15.25 -- 15.40] and
lower reddening [\Ebv\ = 0.09 -- 0.16] than the values adopted here from Harris
(\cite{harris96}).    The use of the distance and reddening of Sosin et al.\
would yield masses about 20\% larger, and luminosities about 0.05 dex larger
than the values given in Table 2.

\subsection{NGC~6121 (M~4)}

Y453 is possibly the hottest  globular cluster star known so far. Other
candidates are three central stars of planetary nebulae (IRAS 18333-2357 in
M~22, Harrington \& Paltoglou  \cite{hapa93}; JaFu1 in Pal~6 and JaFu2 in 
NGC~6441, Jacoby et al., \cite{jamo97}), which however lack model 
atmosphere analyses of their stellar spectra. 
 Such a high \teff\ is not unexpected,  since according to
the Sch\"onberner tracks, a post-AGB star will spend most of its  lifetime at
temperatures greater than 30,000~K. However, as pointed out by  Renzini
(\cite{renz85}), the large bolometric corrections of such hot stars  in the
visible have biased the discovery of post-AGB stars in favour of  cooler 
stars. 

In the \logt --\logg\ plot, Y453 fits well on the 0.546~\Msolar\ post-early 
AGB track of Sch\"onberner (\cite{scho83}).  However, our derived luminosity 
(\logl\ = 2.6) is considerably lower than the Sch\"onberner track at that
\teff\ and \logg, and the derived mass of 0.16~\Msolar\ is astrophysically
implausible. In order to obtain a mass of 0.55~\Msolar, the value of
\logg\ would need to be 5.68 instead of 5.15. This difference is too large to
be accommodated by the spectral fitting, and still would not explain the
discrepancy with the  theoretically expected luminosity.  Therefore, below we
consider some other possible sources of error: \begin{description} \item [Line
Blanketing:] The use of fully line-blanketed NLTE models for  the analysis of
Y453 might result in a somewhat lower temperature  (0.05 dex) without any
changes in surface gravity (Lanz et al. \cite{lanz97}; Haas \cite{haas97}).
Such a lower temperature  would increase the derived mass by about 15\%.

\item [Differential Reddening:]  Cudworth \& Rees (\cite{cud90}) find a 
gradient in the reddening that would increase the adopted reddening for Y453
(\magpt{0}{35})  by about \magpt{0}{015}. Lyons et al. (\cite{lyon95}) report a
patchiness in  the reddening toward M~4 that is at least as significant as the
gradient, and  find a total range of 0.16 mag in \Ebv . An increase by such a
large amount  would still lead to a mass  of only 0.3~\Msolar.  Due to the
non-standard reddening law toward M~4, the reddening correction for Y453 has a
rather high uncertainty in any case.

\item [Distance:] The adopted distance (1.72 kpc) to M~4 is  on the low side of
the range of distance determinations but is supported by both a recent
astrometric measurement (Rees \cite{rees96}), and HST observations of the
main-sequence (Richer et al.\ \cite{rich97}).     Use of the distance given by
Harris (\cite{harris96}; 2200 pc) would give a mass of  0.26~\Msolar. 

\item [Photometry:]   The only ground-based photometry of Y453 of which we are
aware is the photographic photometry  of Cudworth \& Rees (\cite{cud90}), who
also derive a 99\% probability  of cluster membership from its proper motion.
Y453 is among the faintest  stars studied by Cudworth \& Rees, so the
photometric precision might be poorer than their quoted 0.025 mag (which
corresponds to an error of 2\% in M). 

\end{description}

\subsection{NGC~6723}   

The V and B$-$V magnitudes in Table 1 for III-60 are from Menzies
(\cite{menz74}); we are not aware of any other photometry of this star. The 
tabulated photometry for IV-9 is from L.K. Fullton (1997, priv. comm.), who
also gives U--B $= -0.84$. Photometry for IV-9 was also obtained by  Menzies
(\cite{menz74}; V = 14.86, B$-$V $= -0.25$), and  Martins \& Fraquelli
(\cite{mafr87}; V=14.69, B$-$V $= -0.142$).  III-60 and IV-9  fit well on
the 0.546~\Msolar\ post-early AGB track (see above) of  Sch\"onberner, and also
have luminosities (\logl $\sim 3.0$) consistent with  being post-early AGB
stars.

The spectrum of IV-9, however, was difficult to fit with any single model. As
can be seen from  Fig. 1 there is an absorption feature blueward of the
\ion{He}{I} line at  4713~\AA , which matches the \ion{He}{II} absorption line
at  4686~\AA\ in wavelength.  The  \ion{He}{II} line strength and the Balmer
line profiles can be reproduced by a model with \teff\ =   30,000~K, \logg\ =
4.08, and \loghe\ = $-$0.89.  However, this model is inconsistent  with both
the size of the Balmer jump and with the photometric indices  (optical and UV)
of this star.  The photometric data indicate a temperature  of 20000 -- 21000~K
instead. Excluding the absorption feature at 4686~\AA\ from the fit
results in a temperature  of 20700~K and a \logg\ value of 3.34, in good
agreement with the  photometric temperature.  The detection of metal lines
(such as the \ion{O}{II}  absorption lines in BD+33$^\circ$2642 discussed by
Napiwotzki et al., \cite{nahe93}) could  help to decide between the two
temperatures. Simulations with theoretical  spectra, however, show that due to
the low resolution of our data we cannot  expect to see any metal lines. Any
decision will therefore have to await  better data. We keep the cooler
temperature for all  further analysis because of the good agreement with the
photometric indices. 

\subsection{NGC~6752}

B2004 was one of only four post-EHB candidate stars present in the UIT
color-magnitude diagram of  NGC~6752 reported by Landsman et al. 
(\cite{land96}),  and the position of B2004 in the \logt --\logg\ plot (Fig. 3)
is consistent with post-EHB tracks.    Landsman et al.\ estimated \teff\
= 45000~K and \logl\ = 2.12 for B2004 on the basis of IUE spectrophotometry. 
However, the IUE photometry of B2004 had large uncertainties due to the
presence of the nearby ($2.5''$ distant) blue HB star B1995, and the \teff\
(37000~K) and luminosity (\logl\ = 1.94) of B2004 derived  here should be more
accurate. Spectroscopic analyses of the other three post-EHB candidate stars
(B852, B1754, and B4380) in NGC~6752  were presented by Moehler et al.\
(\cite{mohe97}). The four post-EHB stars in NGC~6752 occupy a fairly narrow
range in temperature  (4.5 $< \log$ \teff\ $<$ 4.6) and luminosity ($1.94 < $
\logl\ $ < 2.12$),  and are separated by a large luminosity gap (0.5~dex) from
stars on the populous EHB.    As discussed by Landsman et al.\
(\cite{land96}), these two characteristics are consistent with the 
non-canonical HB models of Sweigart (\cite{sweig97}), which include helium
mixing on the RGB.    However, a more definitive test of EHB evolutionary 
tracks will require a larger sample of post-EHB stars.

\section{Summary}

Among the seven UIT-selected UV-bright stars observed in four  globular
clusters, we find three post-early AGB stars and four post-EHB  stars, but no
genuine post-AGB stars. This can be understood by different  evolutionary
time scales for these stages and by the HB morphology of the  globular clusters
observed: NGC~6752 shows only an almost vertical extended blue horizontal
branch, which means that there is a lack of possible  progenitors for post-AGB
 stars. In NGC~2808 the more luminous UV-bright stars  could not
be observed from the ground due to their proximity to the  cluster center. The
long blue tail of this cluster, on the other hand, provides  plenty of
progenitors for post-EHB stars. M~4 and NGC~6723 lack EHB stars - it  is
therefore no surprise that we do not find any post-EHB stars in these two 
clusters. The blue HB stars found in both clusters are potential progenitors of 
post-early AGB stars. 

We find that post-EHB stars have sub-solar helium abundances, similar  to
their progenitors, which are identified with subdwarf B stars known in the
field of the Milky Way,  while the post-early AGB stars show more or less
solar helium abundances. This difference is not unexpected, because the
diffusive processes at work  on the HB are counteracted by convection and mass
loss on the AGB and  afterwards. 

The present results suggest that post-early AGB stars might be more numerous
than post-AGB stars in globular clusters with a blue HB.     Theoretical
simulations would be useful to determine whether the relative populations of
post-AGB and post-early AGB stars can be accommodated using existing post-HB
evolutionary tracks.     For example, in the models of Dorman et al.
(\cite{dorm93}), post-early AGB stars arise only from HB stars within a narrow
temperature range near $\log$ \teff $\sim$ 4.25.    However, any additional
mass-loss processes, either on the hot HB or during the early AGB, would extend
the \teff\ range of the  progenitors of post-early AGB stars to  cooler HB
stars. Such processes would also reduce the number of PN  candidates and
thereby reduce the discrepancy between the predicted and  observed number of
PNe (Jacoby et al., \cite{jamo97}). Further study of the various globular
cluster  UV-bright stars (post-EHB, post-early AGB, and post-AGB) should
provide  important clues about the mass loss history of post-HB evolutionary
phases. 

\acknowledgements SM acknowledges support by the Alexander von
Humboldt-Foundation, by the director of STScI, Dr. R. Williams through a DDRF
grant, and by the DARA through grant 50 OR 96029-ZA. Thanks go to Michael Lemke
for the many model atmospheres, to Laura K. Fullton for supplying her
photometry prior to publication, to Ben Dorman for  valuable suggestions,
and  to the ESO staff for their support during the observations as well as 
afterwards. This research has made use of the SIMBAD data base, operated at
CDS, Strasbourg, France.

\end{document}